\documentstyle[aps,twocolumn]{revtex}
\begin{document}

\title{Structure and stability of finite gold nanowires}
 
\draft

\author{ G. Bilalbegovi\'c }

\address{Fakult\"at f\"ur Physik, Universit\"at Bielefeld,
D-33615 Bielefeld, Germany}

\date{To be published in Phys. Rev. B, Copyright 1998 by The American Physical
Society}

\maketitle

\begin{abstract}

Finite gold nanowires containing less than $1000$ atoms 
are studied using the molecular
dynamics simulation method and embedded atom potential.
Nanowires with the face-centered cubic structure and
the (111) oriented cross-section are prepared at $T=0$ K.
After annealing and quenching the structure and vibrational properties
of nanowires are studied at room temperature.  
Several of these nanowires form multi-walled  
structures of lasting stability. They 
consist of concentrical cylindrical sheets and resemble multi-walled 
carbon nanotubes. 
Vibrations are investigated by
diagonalization of the dynamical 
matrix. It was found that several percents of
vibrational modes are unstable because of
uncompleted restructuring of initial fcc nanowires. 
    
\end{abstract}

\pacs{68.65.+g,61.46.+w,63.22.+m,81.05.Ys}

Clusters and wires characterized by the nanometer length scales are 
constituents of nanocrystalline materials and components of 
nanoscale electronic devices. These nanostructures show physical 
and chemical properties different from the bulk.
Metallic nanowires are usually fabricated using electron-beam 
lithography \cite{Hegger}. 
The scanning tunneling microscopy experiments recently
gave us a new way of producing nanostructures. 
In these experiments formation
of metallic nanocontacts between the tip and the substrate was observed 
\cite{Pascual}. For such nanowires quantization of conductance was found.
It was shown that similar nanowires are formed between two vibrating
macroscopic wires in contact \cite{Garcia}.
The formation of nanowires and their physical properties, 
such as the electrical conductance, are influenced by elastic, 
structural, and vibrational properties of materials involved.
These properties are also important for applications.
Elastic instabilities of metallic nanocontacts were already studied by
molecular dynamics (MD) simulations 
\cite{Pascual,Stoltze,Sutton,Ruth,Landman,Ciraci,Barnett}.

It is well known that bulk gold crystallizes in
a face-centered cubic lattice.
Several studies have shown that icosahedra are preferred structures for 
small metallic nanoparticles ($< 5 \times 10^3$ atoms) 
\cite{Martin}. Metallic nanowires are not forming 
spontaneously under cluster preparation conditions.
This shows that their cylindrical structure is artificial and
less stable than an icosahedral shape of metallic clusters.
In the laboratories, under special conditions, nanowires of gold and other 
metals are prepared and their cylindrical shapes seem to be rather
stable. Nevertheless, these nanowires should 
transform into more stable structures on a time scale at which 
substantial mass transport could occur. Depending on the number of
atoms, the most stable structure is either a fcc crystal, or
an icosahedral nanoparticle. Therefore,
it is important to analyze the properties of nanowires
on various time and length scales to understand 
the effects of an imposed cylindrical geometry on their
internal structure and stability.
These properties are difficult to study experimentally. 
Recently, special gold nanowires (``nanobridges'') were formed by 
electron-beam irradiation of a gold (001) oriented thin film 
\cite{Kondo}. The structures of these nanowires, which are induced and 
stabilized by the substrate,
were investigated by electron microscopy. The study of the properties
of free nanowires is even more demanding. In contrast, computer 
simulations are suitable for such investigations.

Vibrations of bulk materials are measured by neutron and X-ray
scattering, Raman, and infrared spectroscopy.  
The He scattering and electron 
energy loss spectroscopy (EELS) methods are used to study surface 
vibrations \cite{Phonons}.
Raman scattering is also applied in studies of vibrational properties 
of some clusters, for example of semiconductors \cite{Honea}. 
In addition,
vibrations of argon, water, and ammonia clusters (in the size range
between 25 and 4600 atoms) were recently studied experimentally
by the He scattering technique in a crossed beam arrangement.
\cite{Buck}. Vibrational spectra of semiconductor \cite{Watt} and
carbon \cite{Hiura,Holden} nanowires are measured
by Raman scattering,  but this technique cannot be used for
metallic nanowires because of  specific selection rules.
Vibrational properties of various metallic and non-metallic 
nanoparticles are also investigated theoretically, either
by diagonalization of the dynamical matrix, 
or by calculating the Fourier transform of auto-correlation
functions in a MD simulation
\cite{Kristensen,Garzon,Onida,Wang,Nagy,Jishi,Menon,Adams}.
The computational methods are for now the only techniques that 
can be used for investigations of vibrational properties of metallic 
nanowires.

In this work structural and vibrational properties of  the finite
Au nanowires  
are investigated. Several wires containing from $386$ to $991$ atoms 
are studied at $T=300$ K. 
The MD simulation method based on a well-tested embedded
atom potential is used to produce nanowires. Vibrations are studied by
diagonalization of the dynamical matrix. The tight-binding 
MD method was recently applied in the study of 
finite carbon nanotubes \cite{Menon}. The MD method was
also used for calculation of several properties of infinite wires,
where periodic boundary conditions were applied along the wire axis. 
For example, {\it ab initio} MD simulations of
infinite $Si$ wires were used in 
the studies of optoelectronic properties of porous silicon \cite{Buda}. 
{\it Ab initio} simulations were also applied to analyze
the growth mechanisms for boron-nitride nanotubes \cite{Car}.
Infinite carbon nanotubes were  investigated
by classical and {\it ab initio} simulations \cite{Marco}.
The melting behavior of thin, infinite $Pb$ wires was studied 
by a classical MD simulation \cite{Gulseren}, as well as mechanical 
properties of the $SiSe_2$ wires under strain \cite{Li}.
A recent classical MD
study of Pb and Al ultra-thin infinite wires at $T=0$ K has shown
that several unusual structures appear, for example
icosahedral and helical forms \cite{Erio}. 
In comparison with all these infinite wires,
finite metallic structures studied here represent the smallest
nanowires obtained in the laboratories.
In addition, from theoretical point
of view a finite nanowire is a special version of a cluster whose 
properties deserve an investigation.

The experiments have shown that formation of nanowires is pronounced
for gold \cite{Hegger,Pascual,Garcia}. 
In this work the MD simulation method is used
to prepare gold nanowires. It is well known that 
when classical many-body potentials
of the embedded atom type are used in simulations, a satisfactory 
description of metallic bonding is obtained \cite{Daw}. 
In this simulation a such many-body potential
for gold was employed \cite{Furio}. This potential was chosen because 
of its proven accuracy in various MD simulations for bulk, 
surfaces \cite{Gold}, and clusters \cite{Wanda}.

These simulations started from crystalline nanowires with 
the fcc(111) oriented cross-section at $T=0$ K. 
Nanowires were prepared in an ideal fcc structure by including
all particles whose distance from the nanowire axis is smaller than 
a chosen radius. 
A basic cylindrical MD box consisted of $N_z =12$ layers.
The cross-section with 
the maximal radius of $0.9$ nm was used.  
The total number of moving particles in this MD box was
$386$. Laterally larger nanowires having radii of $1.2$ nm and $1.4$ nm
were also studied, as well as nanowires with $N_z=18$ and 
$N_z=24$ layers.
The prepared ideal samples were first relaxed at $T=0$ K. 
Then MD boxes were heated to $600$ K.
This was followed by a quench to $T=0$ K and heating to $T=300$ K.  
In this procedure no substantial change of the cylindrical shape 
of nanowires 
was observed, either with the temperature, or during time evolution. 
Initially
heating to $600$ K was done, although the proper simulated annealing
and quenching technique used in simulation of metallic 
clusters requires 
higher temperatures ($\sim 0.75$ of the bulk melting temperature, i.e., 
$1000$ K $-1100$ K for gold). 
This was done to prevent melting and collapse into
a drop, but to give to the atoms a possibility to find some local 
minima while keeping a cylindrical shape of a nanowire.
A similar kind of a constrained dynamical evolution of atoms
should also occur for fabricated nanowires. Otherwise,
under real equilibrium conditions depending on the
number of atoms, fcc or icosahedral structures appear.
A time step of $7.14 \times 10^{-15}$ s was used in simulations. Long
runs of $10^6$ time steps (i.e., $7.1$ ns) 
were performed to check a stability of the structures.
It was found that the procedure used for heating and equilibration 
of nanowires gives good results. The structures of lasting stability
were obtained. 
The shapes of these finite nanowires after quenching
were also compared with infinite (111) oriented gold wires with 
similar cross-sections. In these infinite
wires periodic boundary conditions were applied along their axes. 
Apart from slightly rounded ends for smaller
finite wires and more disordering along the vertical direction
for all finite wires,  
no other differences in external shapes of MD boxes
were found at investigated temperatures.

As a result of simulation at $T=300$ K 
several multi-walled cylindrical structures were obtained. 
This can be seen from Figs.~\ref{fig1} and ~\ref{fig2}. Although 
a multi-walled structure exists already
after $10^4$ time steps, to check its 
stability the simulation was carried out up to $10^6$ 
time steps. All multi-walled
structures are preserved on this time scale.
In the cross-sections presented in Figs.~\ref{fig1} (b) 
the hexagonal symmetry of the
fcc(111) surface, which existed in an initial sample at $T=0$ K, 
is replaced by the rings. 
The curved sheets make the concentrical walls of a nanowire.
The central core is surrounded by the
three coaxial cylindrical shells. 
The walls of these shells, i.e., the curved
sheets, are disordered (Fig. ~\ref{fig1} (a)).
Figures ~\ref{fig1} and ~\ref{fig2}  
show that the ``caps'' at the ends of a nanowire are flat 
and that the angle between these flat parts and the lateral wall is nearly $90$
degrees. This is in contrast to the multi-walled carbon nanotubes usually 
capped with semispheres, or polygons \cite{Iijima}. 
However, it was found that the gold
nanowire with the smallest number of atoms has rounded caps and smooth edges
between the cap and the lateral wall. Other bigger gold nanowires possess 
flat ends such as these shown in Figs.~\ref{fig1}  
and ~\ref{fig2}. These flat
caps, as well as the edges between them and lateral walls are disordered on
the atomic scale.
It was found that the formation of multi-walls becomes less pronounced 
when the radius of nanowire increases. For example, in two nanowires with 
$R=0.9$ nm three shells are formed, whereas in a nanowire with $R=1.2$ nm only
two cylindrical walls exist. All nanowires are internally filled.
Such ``the Russian dolls'' arrangement  for carbon nanostructures,
both clusters and wires, was recently the subject of many studies
\cite{Iijima,Ugarte}. 
Multi-walled cylindrical forms were also found for
the layered crystals of $WS_2$ and $MoS_2$ \cite{Tenne}.
In order to check the stability of these structures simulation was actually
carried out up to very long time of $3 \times 10^6$ time steps ($21.3$ ns). 
It was 
found that multi-walled cylindrical structures are preserved. However,
between $10\times 6$ and $2 \times 10^6$ time steps nanowires start 
to rotate about their axes.
For very long simulation time the mass transport within the wire goes in a such
direction that changes its moment of inertia and the angular velocity. The
interplay of rotation and vibrational properties of nanoparticles is rather
involved, as for example shown for vibrations of rotating cluster by Jellinek
and Li \cite{Jellinek}.
The simulation time of $10^6$ time steps ($7.1$ ns) is already much
longer than an average simulation time in other MD studies of clusters.
Therefore, the approach of Ref. \cite{Berry}
where the analysis was stopped before 
noticeable rotation was detectable, is taken. The rotation of nanowires 
certainly also has some structural consequences. Centrifugal forces make the 
wall-structure more pronounced.
Figure \ref{fig3} (a)  shows the only example of nanowire where the shells
do not form, although the fcc(111) cross-section structure 
of an initial sample is lost. 
Therefore, the formation of a multi-walled structure depends on 
the specific size and geometry, i.e., the radius and the length 
for a particular nanowire. The structure
shown in Fig.~\ref{fig3} (a) is cylindrical and
was formed already on a short time scale 
($< 10^4$ time steps) and did not change substantially up to 
$10^6$ time steps.
However, in further simulation  this shape changes
towards an icosahedral structure, as shown in Fig.~\ref{fig3} (b).
The appearance of the outer (111) facets is obvious.
The initial structure of this nanowire was untypical: the diameter 
was slightly ($0.1$ nm) bigger than the length. Therefore,
a such evolution from a cylinder to an icosahedar is not surprising.
Other nanowires with the diameter smaller than the length
are multi-walled and
remain in their cylindrical forms.

Vibrational modes of nanowires were investigated by 
diagonalization of the dynamical matrix. The averaged coordinates of atoms 
obtained in the MD run were taken
as an input. The elements of the force constant matrix for the potential 
used were calculated. Then a such matrix was diagonalized. In this
way the vibrational frequencies were found and from them the density
of states was obtained.
The calculated spectrum was smoothed by convolution
with a Gaussian function.
Vibrational densities of states for two nanowires after $10^6$
time steps are shown in Fig.~\ref{fig4}.
The values of maximal frequencies and the shapes of densities of states
do not change substantially in time. 
The frequency peak at $\sim 3.5$ THz
becomes more pronounced  when the radius of a nanowire increases.   
The maximal frequencies were found to approach $6$ THz.
The phonon dispersion relations for the fcc bulk gold 
were measured by neutron
scattering and the vibrational density of states was calculated using
the fitted force constant models \cite{Lynn}. 
It was found that the maximal frequency is $\sim 4.7$ THz.  
Therefore, the maximal frequencies  calculated here 
for cylindrical multi-walled nanowires 
are higher than for the fcc bulk lattice. 
Figure ~\ref{fig4} (b) shows that the structure of two peaks,
present in the bulk density of states of gold \cite{Lynn} and 
most fcc metals \cite{LB},
appears for gold nanowires containing less than $700$ atoms.
However, the peaks do not have the same size and shape 
as for the fcc bulk gold.
The peaks in Fig.~\ref{fig4} (b) are shifted to lower frequencies in 
comparison with the vibrational density of states of the bulk.
It was found that several percents 
of the total number of vibrational modes  are unstable 
(i.e., for their frequencies is: $\omega^2 < 0$).  
Two examples of a change in the number of unstable modes
are shown in Table ~\ref{table1}.
A similar instability of vibrational modes (so-called soft phonons with
$\omega^2 < 0$) was discussed for surface 
reconstruction \cite{Annalisa} and 
displacive transitions in crystals \cite{Boccara}.
The presence of small number of such modes for nanowires
is a sign of their uncompleted structural evolution.
Similar morphological changes between different 
solid phases on a rapid time scale were recently discovered in small
CdSe nanocrystals \cite{Alivisatos}.

In summary,
MD simulations and lattice dynamical calculations 
for several finite gold nanowires at room temperature 
are presented. The interactions were described by a realistic 
embedded atom potential for gold.
Vibrational properties were investigated by  diagonalization of
the dynamical matrix.   
It was found that cylindrical shapes 
with a multi-walled structure were preserved 
after a long simulation time of  $7.1$ ns.
A nanowire whose initial configuration was such that a 
diameter was slightly bigger than the length evolved
towards an icosahedral shape.
An unusual multi-walled structure of metallic nanowires appears because
of the imposed geometry.
Some preliminary simulations have shown that a multi-walled structure
also exists in infinite (111) oriented gold wires with similar
cross sections. The potential for gold employed in this simulation 
\cite{Furio} was already
used in many studies of surfaces \cite{Gold} and
nanoparticles \cite{Wanda}, and a good agreement with experimental 
results was obtained. Therefore, from a confirmed realibility of the 
potential we should expect that similar 
multi-walled structures exist
in fabricated gold nanowires. In a MD study of Al and Pb
infinite ultra-thin 
(110) and (100) oriented wires at $T=0$ K some two-shell
and three-shell structures were found \cite{Erio}. 
Computer simulations and X-ray reflection studies shown that the liquid-vapor
interface of metals is layered \cite{Rice}. 
The layered structure propagates into the
bulk liquid for a distance of a few atomic diameters. It was found that the
driving force for a such layering is the variation of electron density in the 
surface zone of a metal \cite{Rice}. 
Similar phenomena also give rice to contractive
surface reconstructions. This type of surface reconstruction is very pronounced in gold. The potential for gold employed in this work reproduces correctly both
phenomena: layering at the liquid metal surface \cite{Orio}, 
and reconstruction for all surface orientations \cite{Furio}. 
Therefore, the layering effects present in 
graphite sheets which form carbon nanowires \cite{Iijima}, 
and in the layered crystals of $WS_2$ and $MoS_2$ \cite{Tenne}, 
also exist in gold. Since the layering decreases with the
distance from the surface, it is expected that multi-wall structures disappear 
when the radius of a nanowire is more than a few atomic diameters.
The results of simulation show that a multi-wall structure becomes less 
pronounced when the radius of a nanowire increases.
A jellium model calculation for energetics and quantized conductance of
finite sodium nanowires recently appeared \cite{Yannouleas}. 
A discrete set of magic
wire configurations was found in analogy with the shells in other finite-size
fermionic systems, such as atomic nuclei, metallic and $^{3}He$ clusters. 
Therefore,
the finite size of metallic nanowires increases the tendency for the formation
of multi-walled structures present also because of the layering effect.

All these metallic multi-walled wires, as well as similar structures 
discovered in  experiments for carbon nanowires and nanoparticles 
\cite{Iijima,Ugarte}, and for $WS_2$ and $MoS_2$ \cite{Tenne},
show that such multi-walled morphologies might be quite common 
for materials at nanometer length scales.

I would like to thank M. Ivanda and D. {\v S}estovi{\' c}  
for discussions.

\clearpage

\begin{table}
\caption{Time dependence of
the number of unstable vibrational modes for two nanowires 
(in \% of the total number of modes). 
The nanowire $A$ has the radius $R=1.2$ nm, 
the length $L=2.6$ nm, 
and the number of atoms $n=689$, whereas for the nanowire $B$ 
initially is: 
$R=1.4$ nm, $L=2.7$ nm, and $n=991$. The nanowire $B$ after $10^6$
time steps evolves towards an icosahedral structure 
(see Fig.~\ref{fig3}).
The nanowire $A$ and all other investigated nanowires are
multi-walled.}
\label{table1} 
\begin{tabular}{l l l l l} 
Nanowire$\backslash$Time steps  & $150 \times 10^3$ & $200 \times 10^3$ 
& $500 \times 10^3$  & $10^6$ \\ 
\hline 
$A$  & $1.9$ & $2.1$ & $2.9$ & $1.9$ \\

$B$  & $2.5$ & $3.2$ & $3.9$ & $3.4$ \\
\end{tabular}
\end{table}

\begin{figure}
\caption{Nanowire of $588$ atoms, the length
$4$ nm,
and an average radius of $0.9$ nm,
after a simulation time of $7.1$ ns:
(a) side view,
(b) cross section. 
The trajectory plots refer to a time span of $\sim 7$ ps
and always include the total thickness of the whole MD box.}
\label{fig1}
\end{figure}

\begin{figure}
\caption{
The density plot for a nanowire shown in Fig.~\ref{fig1}. A snapshot 
of morphology is included.}
\label{fig2}
\end{figure}

\begin{figure}
\caption{
(a) Cross section  for a nanowire of radius $1.4$ nm, length $2.7$ nm,
and $991$ atoms, after a simulation time of $7.1$ ns.
In an initial structure of this nanowire the diameter 
is slightly bigger than the length.
(b) Side view of the same sample after $21.3$ ns. 
During simulation a transformation
towards an icosahedral morphology occurs.}
\label{fig3}
\end{figure}

\begin{figure}
\caption{
Vibrational density of states for nanowires of radius $R$, length $L$,
and number of atoms $n$:
(a) $R=0.9$ nm, $L=5.4$ nm, $n=784$,
(b) $R=1.2$ nm, $L=2.6$ nm, $n=689$.}
\label{fig4}
\end{figure}

\end{document}